\documentclass{article}

\usepackage{arxiv}
\usepackage{amsmath}
\usepackage[utf8]{inputenc} 
\usepackage[T1]{fontenc}    
\usepackage{hyperref}       
\usepackage{url}            
\usepackage{booktabs}       
\usepackage{amsfonts}       
\usepackage{nicefrac}       
\usepackage{microtype}      
\usepackage{lipsum}
\usepackage{graphicx}
\graphicspath{ {./images/} }
\usepackage[version=4]{mhchem}

\title{Machine Learning Refinements to Metallicity-Dependent Isotopic Abundances}

\author{
Haoxuan Sun \\
  Department of Physics \& Astronomy\\
  Macalester College\\
  Saint Paul, MN 55105 \\
  \texttt{hsun@macalester.edu} \\
}

\begin{document}
\maketitle
\begin{abstract}
The project aims to use machine learning algorithms to fit the free parameters of an isotopic scaling model to elemental observations. The processes considered are massive star nucleosynthesis, Type Ia SNe, the s-process, the r-process, and p-isotope production. The analysis on the successful fits seeks to minimize the reduced chi squared between the model and the data. Based upon the successful refinement of the isotopic parameterized scaling model, a table providing the 287 stable isotopic abundances as a function of metallicity, separated into astrophysical processes, is useful for identifying the chemical history of them.  The table provides a complete averaged chemical history for the Galaxy, subject to the underlying model constraints. 
\end{abstract}


\section{Introduction}
The yields of stellar simulations are dependent on the star's initial isotopic composition. During hydrostatic burning phases, the initial composition is crucial for neutron capture reactions on initial metals, which affects the abundance of odd-z nuclei(\cite{pignatari2010weak}). The detailed stellar abundances influence the opacity of the star, which in turn influences the star's structure as well as the loss of mass and angular momentum, thereby altering the late stellar evolution. Understanding $\gamma$-process abundances, which use s- and r-process isotopes as seeds, requires information on the initial abundances of heavy isotopes (\cite{rauscher2002nucleosynthesis} \& \cite{arnould2003p}). The objective of galactic chemical evolution (GCE) is to comprehend how the abundances of the elements and their isotopes changed from the big bang to the present day, and it can be used to obtain the isotopic abundances at any metallicity for use as inputs for stellar simulations. Traditional GCE models typically require nucleosynthesis yields from stellar simulations as inputs, but the problem with this approach is that in order to provide self-consistent nucleosynthesis yields, the stellar simulations require a full initial set of isotopic abundances (\cite{timmes1995galactic};\cite{chiappini1997chemical};\cite{costa2009chemical}). The construction of an astrophysical model of all stable isotopes, based on physical principles for production sites and mechanisms, is a complementary strategy to conventional GCE methods. The completed model then provides the Galaxy's average isotopic history, subject to the employed approximations (\cite{kobayashi2006galactic}). Compared to full GCE calculations, this method is comparatively simple and approximative, but it improves upon the standard of scaling isotopic solar abundances by a constant factor.

The study of chemical abundance in physics is an important aspect of understanding the composition of matter in the universe. Chemical abundance refers to the relative amounts of different chemical elements present in a given sample. This information is crucial for understanding the formation and evolution of stars, galaxies, and the universe as a whole. All stellar evolution models for nucleosynthesis necessitate a beginning point for isotopic abundance (\cite{portinari1998galactic}). Except for the Sun, our understanding of the isotopic abundances of stars is generally incomplete. We offer parameters for models of a complete average isotopic decomposition as a function of metallicity, which are fitted to observational data, as opposed to the conventional forward galactic chemical evolution modeling that incorporates star yields beginning with big bang nucleosynthesis. This method of machine learning uses the grid search algorithm to identify the parameter values with the lowest reduced chi-square value, resulting in a better fit of the model to these real data. Our fittings of parameter finds the light elements with great accuracy, within a $8.7\%$ uncertainty compared with the original results but as well as a better performance in the heavier element corresponding to strong $s-$process and main $s-$process. 

\section{Overview}
\label{overview}

This project primarily focuses on two main components aimed at improving the astrophysical model completed by (\cite{west2013metallicity}). The research employs the application of machine learning algorithms to refine the model, using the grid search and random forest algorithms, and analyzing the impact of these adjustments on isotope abundances.

The background context of the problem lies in the need for an improved astrophysical model that can accurately predict isotope abundances in stars. This is crucial for understanding stellar nucleosynthesis processes and the evolution of elements in the universe. The general solution provided by this project involves the integration of machine learning techniques to optimize parameters within the model, for the specific purpose of improving stellar simulation inputs.

The grid search algorithm primarily optimizes most reaction parameters, with the exception of the strong s-process. Conversely, the random forest algorithm focuses on adjusting the scaling and shift factors of the s-process. We evaluate our current progress by comparing it both numerically and visually to previous work, utilizing Python-based code.

After successfully refining the model, we compute the abundances of all 287 isotopes for various reactions, considering different initial stellar metallicities. The reactions are color-coded and stacked together to generate a comprehensive abundance graph, which is further normalized using solar isotope abundances.

Subsequently, we select specific isotopes for in-depth analysis and discuss the overall trends observed. This holistic approach allows us to gain a deeper understanding of the underlying astrophysical processes and evaluate the effectiveness of the machine learning algorithms in improving the model.

\section{Astrophysical Processes}
The origin of the elements in the Universe is a fundamental question in astrophysics. Understanding the processes responsible for the production of isotopes is crucial for gaining insights into the galactic chemical evolution models used in nucleosynthesis studies. Thus, various astrophysical processes that contribute to the formation of isotopes, including hydrostatic burning, supernovae, the $r$-process, and the $s$-process will be discussed.

Primary processes are mainly producing metals directly from H and He. Nuclear fusion is the primary process by which atomic nuclei combine to form more massive nuclei, releasing energy in the process (\cite{B2FH}). This process powers the energy production in stars, including our sun, and is responsible for the production of heavier elements \cite{bethe1939energy}. While Type Ia supernova event is a powerful and luminous explosion of a massive star at the end of its lifecycle. It is one of the most energetic events in the universe and results in the dispersion of elements and the formation of neutron stars or black holes (\cite{woosley2005physics}).

Neutrino-induced nucleosynthesis, also known as neutrino nucleosynthesis, is a process in which neutrinos interact with atomic nuclei, resulting in the production of new elements. This process is particularly important in core-collapse supernovae, where a large number of neutrinos are produced, and their interactions contribute to the synthesis of some isotopes (\cite{takahashi1994nucleosynthesis}).

The r-process is a nucleosynthesis process that occurs in explosive astrophysical events, such as core-collapse supernovae or neutron star mergers. It involves the rapid capture of neutrons by seed nuclei, leading to the production of heavy, neutron-rich isotopes, including many of the elements heavier than iron. The R-process is responsible for the synthesis of about half of the elements heavier than iron (\cite{cowan1991r}).

The $\nu p$-process, or neutrino-p process, is a nucleosynthesis process that occurs in the neutrino-driven winds from protoneutron stars during a core-collapse supernova. It is driven by the interaction of neutrinos with protons, leading to the production of elements heavier than iron via rapid proton captures on existing seed nuclei. The $\nu p$-process is thought to be responsible for the synthesis of some neutron-deficient isotopes of heavy elements (\cite{frohlich2006neutrino}).

Secondary processes are different from primary processes since they make metals from pre-existing metals made earlier in a primary process and could be divided into three major types. The weak s-process occurs in massive stars (with masses greater than 10 solar masses) during core helium and shell carbon burning. It synthesizes elements with atomic numbers between iron and strontium (Z = 38), contributing to the production of elements in the lower mass range of the s-process. The main s-process occurs in low-to-intermediate-mass stars (with masses between 1 and 8 solar masses) during the asymptotic giant branch (AGB) phase. It is responsible for producing elements with atomic numbers between strontium (Z = 38) and bismuth (Z = 83). The main s-process is responsible for synthesizing most of the s-process elements in the universe. The strong s-process is a hypothetical nucleosynthesis process that would occur in an environment with an extremely high neutron density. It is not currently associated with any known stellar environment (\cite{clayton1983principles}).

\section{Model Description}
In this project, we mainly utilize the existing model from \cite{west2013metallicity} and briefly discuss the models that we fit in the machine learning algorithm. Generally, isotopes can be divided into light and heavy isotopes, with light isotopes having masses lower than $56$, which is the threshold of ${^{56}Fe}$, and heavy isotopes having masses greater than this threshold. Light isotopes are primarily formed through BBN, Type Ia supernova, or massive star reactions. We make assumptions on the Type Ia supernova model using the equation $f\dot X_{56}^\odot/X_{56}^{Ia}$, where $X_{56}^{Ia}$ is the solar abundance of ${^{56}Fe}$, and $X_{56}^{Ia}$ is a scaling factor. $f$ is a free parameter in the model, determined by fitting the elemental scalings against available data. We assume from \cite{west2013metallicity} that massive star yields were scaled to the remaining contribution to solar, resulting in the scaling model for Type Ia supernova as $(1-f)\cdot X_{56}^\odot/X_{56}^{Ia}$. This model, however, is only for the specific isotope ${^{56}Fe}$; we further assume that for all other isotopes between $12\leq A\leq 56$, additional scaling preserves the ratio of each isotopic contribution for massive star yields.

\begin{equation}
\begin{aligned}
X_{i,f}^{massive}&=(\frac{X_i^\odot}{X_{i,0}^{massive}+X_{i,0}^{Ia}})X_{i,0}^{massive}\\
X_{i,f}^{Ia}&=(\frac{X_i^\odot}{X_{i,0}^{massive}+X_{i,0}^{Ia}})X_{i,0}^{Ia}
\end{aligned}
\end{equation}

In this equation, $X_{i, 0}$ and $X_{i, \mathrm{f}}$ represent the original and fitted abundances of isotope $i$ for massive or Type Ia contributions, denoted by superscripts, and the solar abundance of isotope $i$ is given by $X_i^{\odot}$.

After considering all model assumptions, we combine them into a more general scaling model. To create a linear relationship for the scaling model, we apply logarithmic methods to their respective abundances given by the normalized Population III simulation. For each isotope, the abundances are scaled according to:

\begin{equation}
\begin{aligned}
\log \left(X_i^*(\xi)\right)=m_i\left(\log (\xi)-\log \left(\xi_{\text {low }}\right)\right)+\log \left(X_i^{\text {sim }}\right)
\end{aligned}
\end{equation}
where $X_i^*(\xi)$ is the massive abundance of isotope $i$ as a function of the model parameter, and $\log \left(\xi_{\text {low }}\right)=-2.5$. The slope is defined as, \\$m_i \equiv\left(\log \left(X_{i, f}^{\text {massive }}\right)-\log \left(X_i^{\text {sim }}\right)\right) /\left(\log \left(\xi_{\odot}\right)-\log \left(\xi_{\text {low }}\right)\right)$, where $X_{i, f}^{\text {massive }}$ is the massive star contribution to the solar abundance, and obviously $\log \left(\xi_{\odot}\right)=0$. Though, we finish the modeling of massive star reaction for light isotope, type Ia supernova reaction occured in creating light isotope is not negligible. One specific form which satisfies all three constraints discussed is given, e.g., by a scaled and shifted hyperbolic tangent base function,
\begin{equation}
  \begin{aligned}
X_i^{\mathrm{Ia}}(\xi)= & X_{i, \odot}^{\mathrm{Ia}} \cdot \xi \cdot[\tanh (a \cdot \xi-b)+\tanh (b)] /[\tanh (a-b) \\
& +\tanh (b)] 
  \end{aligned}
\end{equation}
Where $X_i^{Ia}(\xi)$ is the supernova abundance of isotope $i$ as a function of the model parameter. The specific value for Type Ia onset is determined by fitting the free parameters, $a$ and $b$, against available data. Note the hyperbolic tangent function is tempered with a linear factor of $\xi$ to ensure the behavior is linear near solar.After discussing the assumption made for the light isotopes, heavy isotopes involve more complicated models that need to be apply. Thus we are going to conclude the existence for Neutron Capture and p-isotopes. This reaction will take place in most of the heavier isotopes. And this Neutron Capture and p-isotopes include the main $s$-process $(l s, h s$, and "strong" component), weak $s$-process, $r$-process, $v p$-process, and $\gamma$-process contributions are parameterized as power laws. In the equation, $X_i^{\odot}$ is the solar abundance of the isotope i, where $X_i^*(\xi)$ with different subscripts of $strong, ls, hs, ws, r, \mu p, \gamma$ is the respective processes' abundance of isotope i as a function
of the model parameter.

\begin{equation}
 \begin{gathered}
X_i^{\mathrm{strong}}(\xi)=c\left[1-\frac{\tanh (d \cdot \xi+g)}{\tanh (d+g)}\right]+X_{i, \odot}^{\mathrm{strong}} \\
X_i^{\mathrm{ls}}(\xi)=X_{i, \odot}^{\mathrm{ls}} \cdot \xi^l \\
X_i^{\mathrm{hs}}(\xi)=X_{i, \odot}^{\mathrm{hs}} \cdot \xi^h \\
X_i^{\mathrm{ws}}(\xi)=X_{i, \odot}^{\mathrm{ws}} \cdot \xi^w \\
X_i^r(\xi)=X_{i, \odot}^r \cdot \xi^p \\
X_i^{v p}(\xi)=X_{i, \odot}^{v p} \cdot \xi^p \\
X_i^\gamma(\xi)=X_{i, \odot}^\gamma \cdot \xi^{\frac{h+p}{2}+1} 
  \end{gathered}
\end{equation}

While there are few isotopes undergoes the reaction Hydrogen Burning, Classical Novae, v-process, and Galactic Cosmic Ray Spallation, but will be scaled in an subtraction of mass fraction of the helium isotopes and mass fraction of the helium isotopes from the solar value of total metallicity. The remaining isotope of hydrogen, ${ }^1 \mathrm{H}$, was scaled according to,

\begin{equation}
  \begin{aligned}
X_i^{\mathrm{H}}(\xi)=X_{i, \odot}^{\mathrm{H}} \cdot\left[1.0-\xi \cdot Z_{\odot}-Y(\xi)-D(\xi)\right]
  \end{aligned}
\end{equation}
where $Y(\xi)$ is the mass fraction of the helium isotopes, $D(\xi)$ s the mass fraction of deuterium, and $Z_{\odot}$ is the solar value of otal metallicity. 

Since all the scaling models are introduced for all these 287 different isotopes, it is confident to say that there are complicity when with in the same element, for example, Iron, could involve different reactions since $56$ isomass is the threshold for between heavy and light isotopes. For ${Fe}$, the reaction will be 
\begin{equation}
\begin{aligned}
&\begin{aligned}
{ }^{56} \mathrm{Fe}(\xi)= & { }^{56} \mathrm{Fe}_{\odot}^{\mathrm{Ia}} \cdot \xi \cdot[\tanh (a \cdot \xi-b)+\tanh (b)] /[\tanh (a-b) \\
& +\tanh (b)]+\mathrm{Fe}_{\odot}^* \cdot 10^{m_{\mathrm{Fe} 56} \cdot\left(\log (\xi)-\log \left(\xi_{\mathrm{low}}\right)\right)+\log \left(X_{\mathrm{Fe} 56}^{\mathrm{sim}}\right)}
\end{aligned}\\
&{ }^{57} \mathrm{Fe}(\xi)={ }^{57} \mathrm{Fe}_{\odot}^{\mathrm{ws}} \cdot \xi^w+{ }^{57} \mathrm{Fe}_{\odot}^* \cdot 10^{m_{\mathrm{Fe} 57} \cdot\left(\log (\xi)-\log \left(\xi_{\mathrm{low}}\right)\right)+\log \left(X_{\mathrm{Fe} 57}^{\mathrm{sim}}\right)}
\end{aligned}
\end{equation}
Thus a machine learning algorithm is thus to be used under these situation since we are dealing with the models that could change gradually as the parameters are changing which in this case are interconnected. 
\section{Methodology of Work}
In the published study under discussion, the authors developed a novel algorithm for constructing elemental abundance ratios \([X/\text{Fe}]\) using isotopic scaling functions and fitting them to observational data. By systematically analyzing isotopic contributions to the solar abundances of various elements, they derived elemental scaling relations that account for multiple nucleosynthetic processes. Their methodology not only provided a robust way to evaluate elemental ratios \([X/\text{Fe}]\) and \([\text{Fe}/\text{H}]\) as functions of the continuous technical parameter \(\xi\), but also enabled efficient comparison with observational data through \(\chi^2\) analysis.

Compared to the published work, the previous model has an over-reliance on Mg data to fit free parameters. In this project, I employ machine learning techniques to simultaneously fit all the elemental abundance data along with all the free parameters, thereby addressing the limitations of the previous model and offering a more comprehensive solution. 

Similarly, the treatment of the strong \(s\)-process in the previous work involved crudely fitting the strong \(s\)-process parameter by hand using Pb data. In contrast, we consider more data points for this typical astrophysical process and develop a better solution to find parameters. This project aims to improve the understanding of nucleosynthesis and the chemical evolution of the universe, making it a suitable topic for an honors thesis in astrophysics.

the observational data analysis was conducted through a rigorous and systematic process. The data points, which represent various measurements of chemical abundances of elements in stars, were analyzed in the context of [X/Fe] versus [Fe/H] space. This space is used to study the chemical evolution of galaxies and stellar populations, where [X/Fe] denotes the abundance ratio of element X to iron (Fe), while [Fe/H] signifies the abundance of iron relative to hydrogen (H).

To account for uncertainties in the measurements, a Gaussian spread was assigned to each observational data point, based on a nominal uncertainty of 0.15 dex. The Gaussian spread is a probability density function, which describes the distribution of the data points around their true values. It is mathematically represented as:
\begin{equation}
\begin{aligned}
f_i=\exp \left\{-0.5 \cdot\left(\left[\left(x_i-x_0\right) / \sigma_x\right]^2+\left[\left(y_i-y_0\right) / \sigma_y\right]^2\right)\right\}
\end{aligned}
\end{equation}

Here, $f_i$ represents the Gaussian contribution of the $i^{th}$ data point, $(x_i, y_i)$ are the coordinates of the data point in the [X/Fe] vs [Fe/H] space, and $(x_0, y_0)$ are the true values of the data point. The parameters $\sigma_x$ and $\sigma_y$ represent the standard deviations of the Gaussian spreads in the $x$ and $y$ directions, respectively.

The [X/Fe] vs [Fe/H] space was then divided into a series of bins, which facilitated the organization and analysis of the data. Within each bin, the Gaussian contributions of all the data points were used to compute an average and standard deviation. This process helped to mitigate the effects of uncertainties in individual data points and provided a more reliable basis for further analysis.

\section{Machine Learning}
Methods with machine learning involve supervised machine learning and unsupervised machine learning, where the first one has variables given a certain target and the other one operates upon only the input data without outputs or target variables. Since previous work proves that there has already been successful fitting on most of the models, we then choose the supervised machine learning method to further improve the performance of the model. Here, we introduced a method of grid search, where we input the range of the variables and enable the machine to get the sufficient time to run through all possible combinations within a line-space of 0.01. Though, this machine learning algorithm is time intensive while also need strong computation ability for computers, still this is the best method that could result in best parameter fittings for our models. This grid search method, scans over $7^{10}$ possible parameters given the existing data. By using this typical machine learning method, we could cover all the processes that potentially correlate between isotopes. The test success will be based on the least reduced chi-square values. Previous work involves a method that assigns Gaussians to data points based on uncertainties and divides the region into grids. However, the exact correlation is unsupported by data, and the model's assumption of scaling may not be statistically accurate at very low metallicities due to insufficient data and the influence of individual astrophysical events. Thus the way that we come out to test whether model fits or not is the reduced chi-square method. This method states the following equation (\cite{bevington2003data}):

\begin{equation}
\begin{aligned}
x_r^2=\frac{1}{N}\sum_{i=1}^{N}(\frac{x_i-\mu_i}{\sigma_i})^2
\end{aligned}
\end{equation}
Within reduce chi square equation, N represents the number of bins we choose specifically, while $\mu_i$ is the mean of the data $\sigma_i$ be the standard deviation and $x_i$ symbols the model use. In case we do not lose the generality, we use the bin size of 200 for all isotopes fitting with standard deviation value of 0.1. The bin size means that we cut the whole graph with 200 sections but we ignore the initial sections of line because there was some defects how the data is constructed. While we set each bin size, we use the model data compared with the observed data and get one bin size reduced chi-square and sum them up at last. As long as we got all the chi-square, according to (paper), reduced chi-square will behave linearly in the statistics such that we could sum all the isotopes reduced chi-square values all together to get a final reduced chi-square value. In other words, the least reduced chi-square corresponds a best parameter set for the result. 

\section{Grid Search}

Grid search is a parameter optimization technique that searches over a specified set of hyperparameter values to find the optimal combination for a given model (\cite{bergstra2012random}). In general, since the previous work fitted free parameters well for most light isotopes, as evidenced by the frequent use of Magnesium isotope data in his model, we set the range of parameters within $0.5$ of the initial parameter values. For example, if the original parameter for a massive star reaction is $2.5$, we instruct the machine to search within the range of $2$ to $3$, outputting a reduced chi-square value for each parameter value examined. The step size is set to $0.05$, meaning that for each iteration of parameter search, there is a $0.05$ increase in parameter values and a corresponding reduced chi-square value output. This logic is applied to all reaction parameters in the model, generating a large dataset of reduced chi-square values. We calculate the linear sum of the reduced chi-square values for all parameter sets and identify the least value as the best fit set of parameter values. Although grid search is a straightforward way to determine the best fit parameter values, it is computationally intensive and time-consuming. As a result, we explore a new algorithm for the second part of our study, focusing on the strong $s-$process.

\section{Random Forest}

Random Forest is an ensemble learning method that builds multiple decision trees and combines their predictions to improve accuracy and reduce overfitting (\cite{breiman2001random}). As we have a numerical method to assess the effectiveness of our machine learning approach, we conduct preliminary result estimations before implementing the machine learning method. We focus on typical reactions occurring in heavy elements, such as Lead, and believe that the previous parameter values for these reactions are not representative of the true values. Consequently, we closely examine the machine learning method for these elements and introduce the Random Forest algorithm to better understand how machine learning works for these specific reactions.

Using the Random Forest algorithm significantly reduces the time required to compute ideal parameter values while maintaining the same logic as the reduced chi-square method. We instruct the computer to follow the trend of reducing reduced chi-square values when adjusting parameter values. If the computer identifies an increasing trend in reduced chi-square values, it will change the parameter values to reverse the trend and reduce the reduced chi-square value. Once a minimum reduced chi-square value is reached, the computer will continue searching for more parameter values and corresponding reduced chi-square values unless a peak value has been found, at which point it will output the results. This algorithm is specifically used for determining the values of the strong $s-$process, as there are limited datasets to train on and only a few isotopes undergo this process for reactions. The best training results are achieved for the Lead isotope, which will be discussed in the analysis section.

\section{Analysis on Grid Search Algorithm}
In this honor thesis, we aim to provide a more quantitative analysis of a machine learning algorithm specifically designed for comparing astrophysical models based on their parameter values. This innovative algorithm leverages a reduced chi-square calculation to generate numerical values for each pair of parameters, thereby enabling direct comparison of these values to determine the best-fit model.

To assess the performance of our algorithm, we concentrate on astrophysical processes, excluding strong $s$-processes, which are key to understanding the nucleosynthesis of heavy elements in stars. We implement a grid search algorithm and systematically evaluate the contributions of various isotopes to the total reduced chi-square values, thus offering a more comprehensive understanding of the model's performance.

For instance, Magnesium isotopes demonstrate the lowest contribution to the total reduced chi-square value, with a value of $0.78$. This finding supports the methodology of previous work, which heavily relied on Magnesium data to fit free parameters due to its reliable and abundant data. In stark contrast, Germanium isotopes display the highest contribution to the reduced chi-square value, at $46.9$. This significant discrepancy can be primarily attributed to a lack of data points and a large inconsistency in the available data for Germanium isotopes.

\begin{figure}[!htb]
  \centering
  \includegraphics[width=0.8\textwidth]{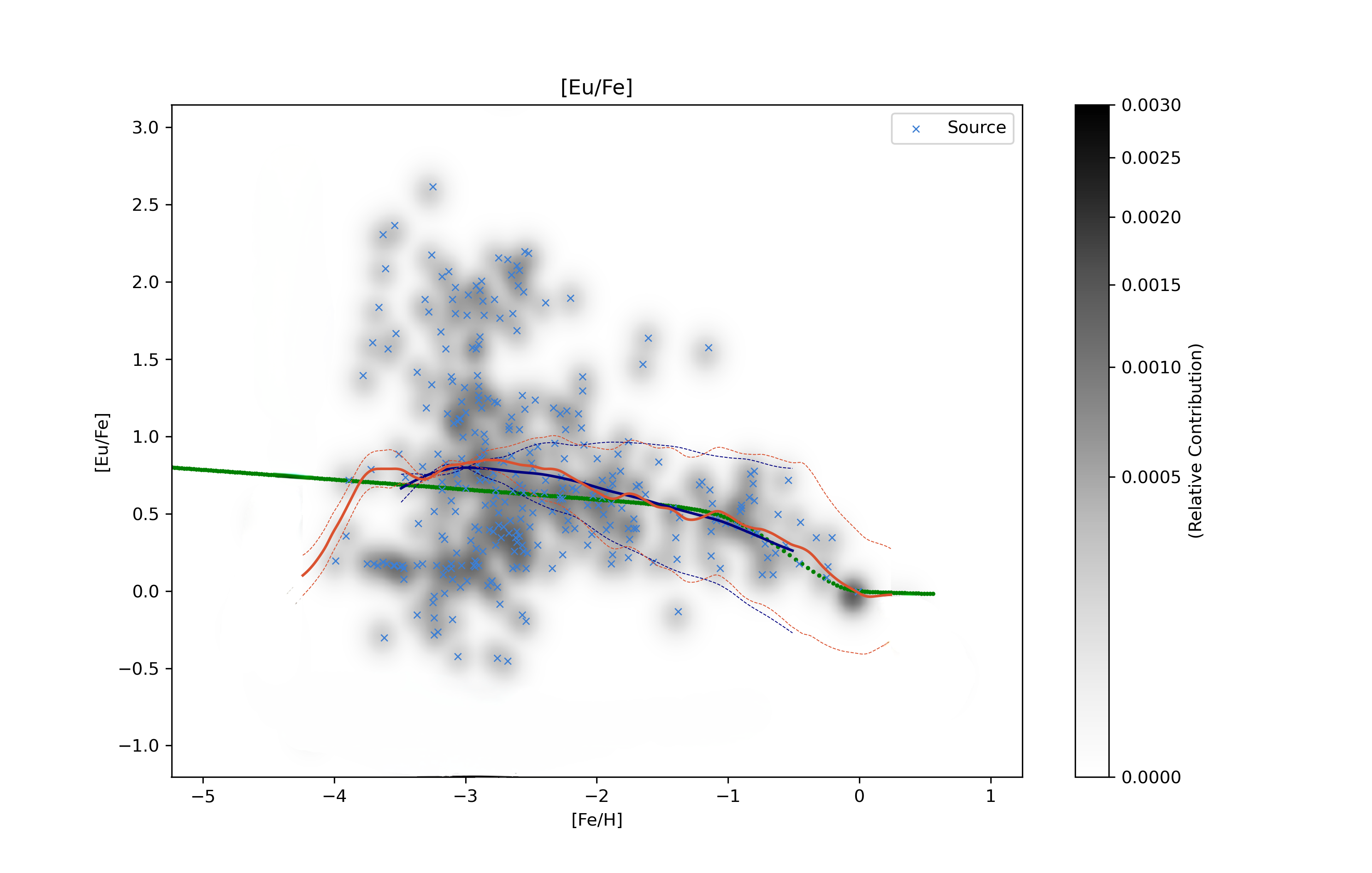}
  \caption[Published Results for {[Eu/Fe]} from \cite{west2013metallicity}]{Published Results for [Eu/Fe] from \cite{west2013metallicity}, the green line represents the theoretical model curve, while the blue line signifies the observed average data trend. The black shaded region encapsulates the uncertainty associated with individual data points, which are further indicated by blue cross symbols throughout the graph.}
  \label{fig_3.1}
\end{figure}

\begin{figure}[!htb]
\centering
\includegraphics[width=0.8\textwidth]{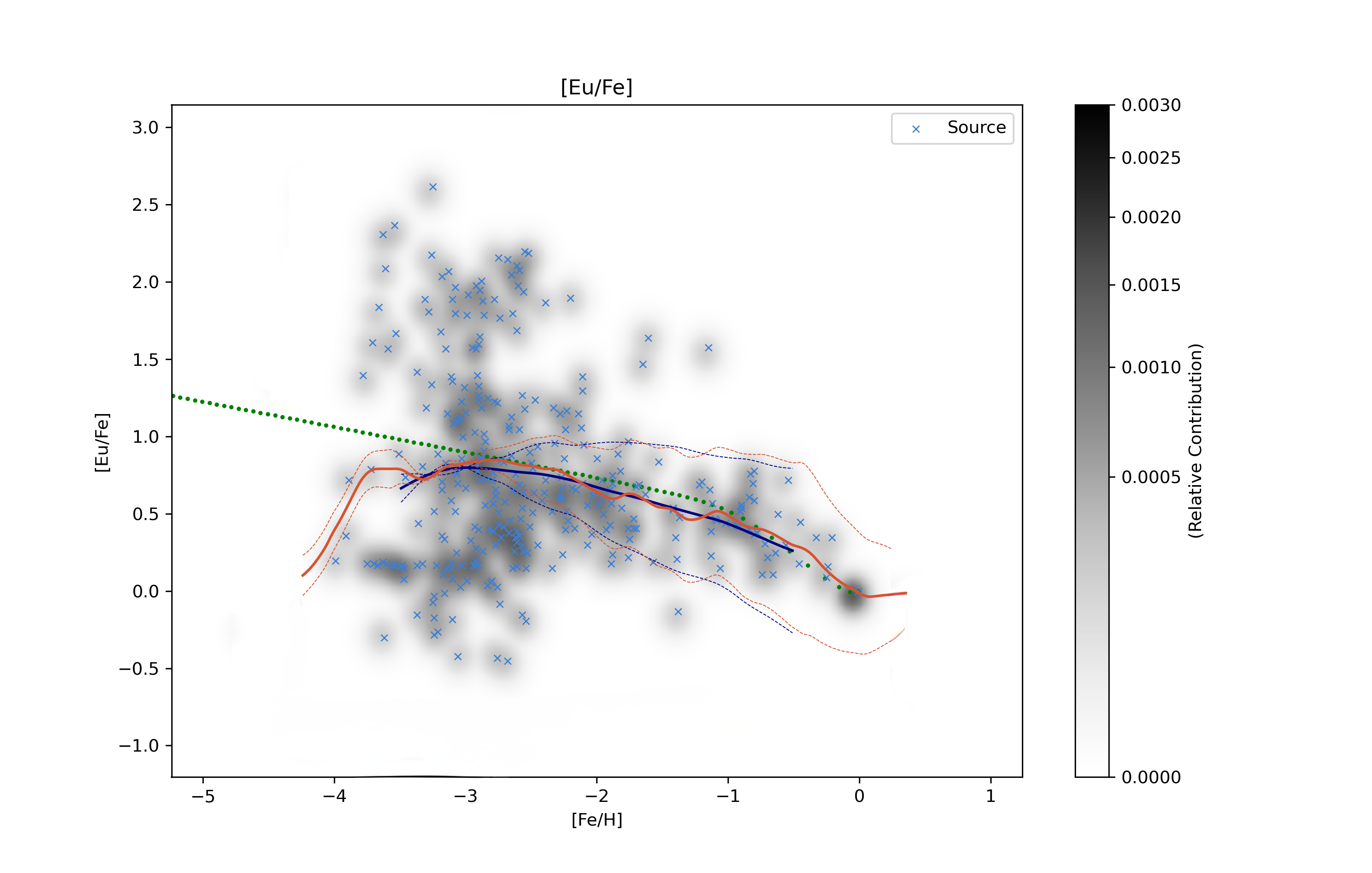}
\caption[The resulting model for{[Eu/Fe]} found by parameter fitting]{The resulting model for [Eu/Fe] found by parameter fitting. Displaying the green line as the model prediction, the blue line as the average data trend, the black shaded area representing uncertainty in individual data points, and blue crosses as particular data points.}
\label{fig_3.2}
\end{figure}

To further validate our findings, we conducted a visual comparison of graphs generated using two distinct sets of parameters. Figure \ref{fig_3.1} presents the published [Eu/Fe] results from West $\&$ Heger (2013), and Figure \ref{fig_3.2} showcases the resulting model for [Eu/Fe] found by parameter fitting. The graphs employ a green line for the model prediction, a blue line for the average data trend, a black shaded area to convey uncertainty in individual data points, and blue crosses to indicate specific data points, providing an in-depth visual comparison of the astrophysical findings. As seen in Figures \ref{fig_3.1} and \ref{fig_3.2}, the salient difference is a better fit against the data averages using the machine learning method. This better fit comes at the expense of having a slightly non-zero slope to the model line at low metallicities, which makes less physical sense -- since at low  metallicities, both Eu and Fe are made in primary processes, hence their ratio should be a flat function of metallicity. Nevertheless, we recognize that there is no available data to constrain the model at such low metallicities, where the chemical evolution may have been more stochastic and without a well defined averaged anyway. 

\begin{figure}[!htb]
  \centering
  \includegraphics[width=0.8\textwidth]{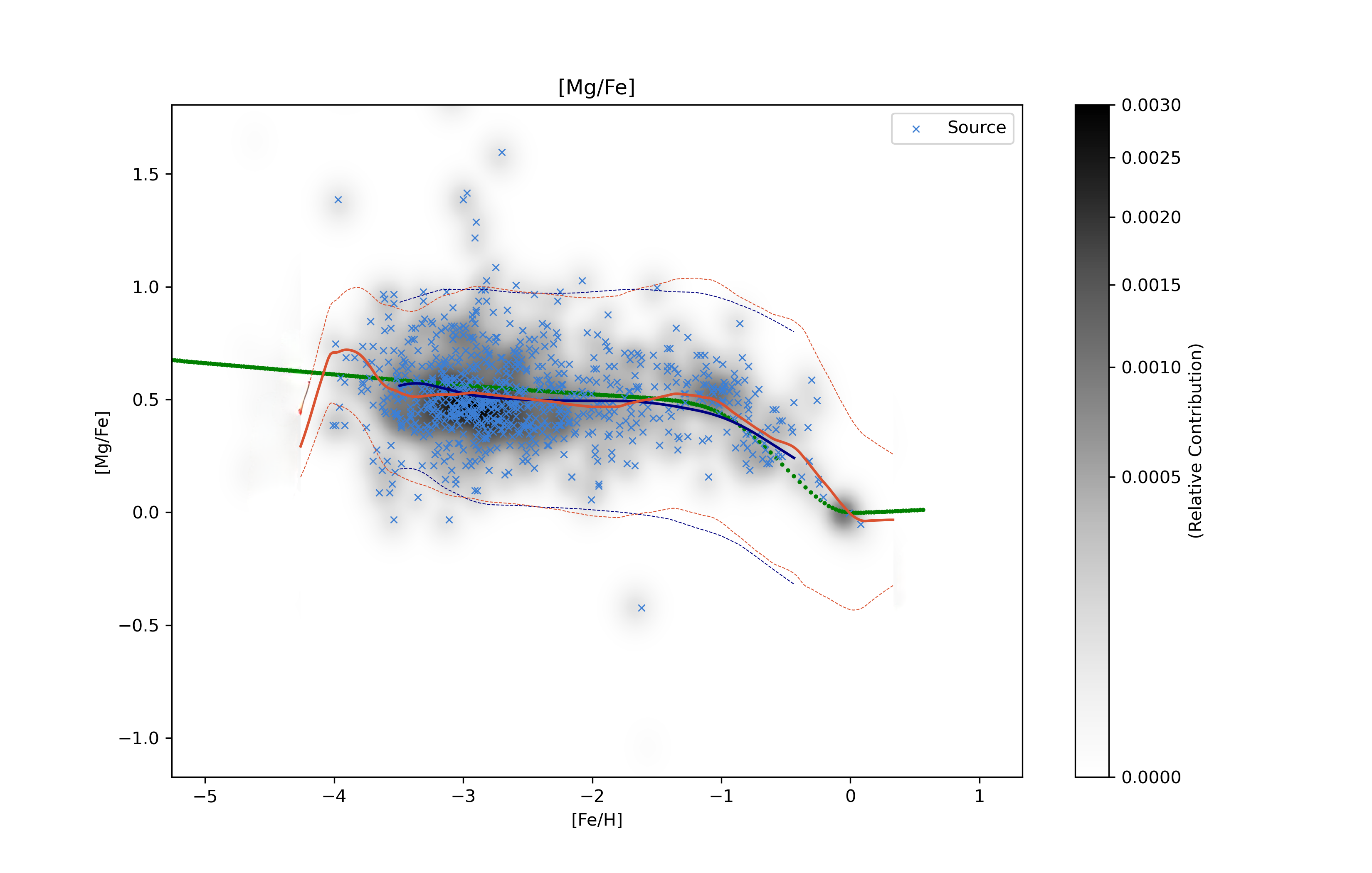}
  \caption[Published Results for {[Mg/Fe]} from\cite{west2013metallicity})]{Published Results for [Mg/Fe] from \cite{west2013metallicity}. The green line symbolizes the predicted model curve, whereas the blue line reflects the mean data line. Uncertainty in each data point is showcased by the black shaded zone, with blue crosses marking individual data points.}
  \label{fig_3.3}
\end{figure}

\begin{figure}[!htb]
  \centering
  \includegraphics[width=0.8\textwidth]{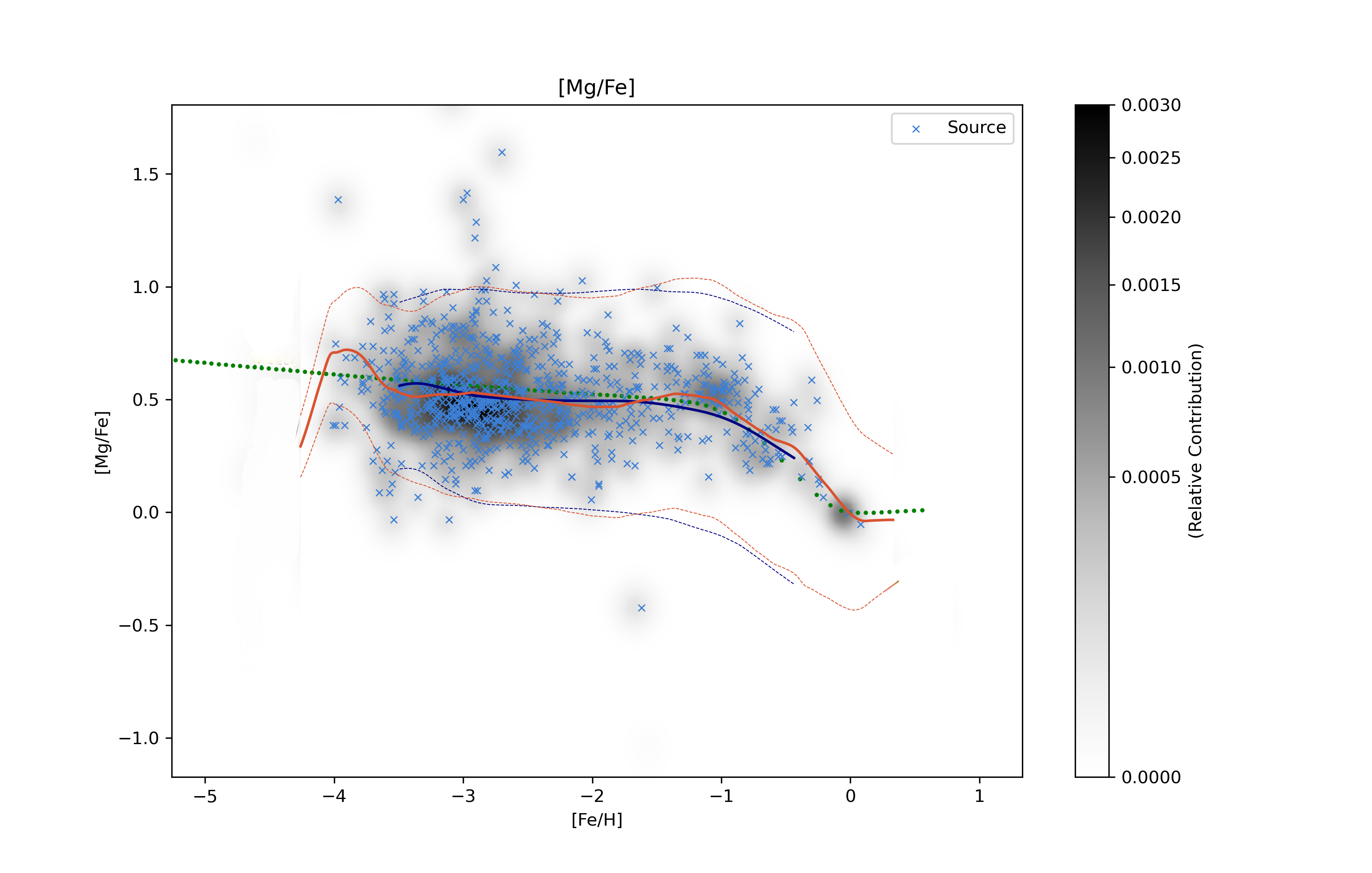}
  \caption[The resulting model for {[Mg/Fe]} found by parameter fitting]{The resulting model for [Mg/Fe] found by parameter fitting.The green line denotes the model's projected outcomes, while the blue line corresponds to the average data trajectory. Individual data point uncertainty is demonstrated by the black shaded area, and blue crosses pinpoint specific data points.}
  \label{fig_3.4}
\end{figure}

The visual comparison reveals minimal changes in Magnesium isotopes, as expected, considering the previous work relied on Magnesium isotopes to determine several best-fit values due to the abundance of data. Figure \ref{fig_3.3} combines the published [Mg/Fe] results from West $\&$ Heger (2013), and Figure \ref{fig_3.4} displays the resulting model found by machine learning parameter fitting. Both figures effectively illustrate the predicted model curve and the mean data line, as well as the uncertainty in each data point represented by the black shaded zone. Although the present model does not offer much improvement over the previous work for this element, it does serve as a consistency check demonstrating that the machine learning process reproduces what the previous work did quite well.

\section{Analysis on Random Forest Algorithm}
In this section, we conducted a thorough analysis of the success of the random forest algorithm in refining our astrophysical model. As mentioned in previous sections, this algorithm is specifically employed in the region where the $s$-process reactions occur. Due to this assumption, we have a limited data-set available for training; however, our algorithm performed remarkably well for isotopes with few data points provided.

For initial parameter values of $d=150$ and $g=-0.23$, the machine learning algorithm successfully modified these values to $d=250$ and $g=-0.05$. This adjustment effectively reduced the reduced chi-square value from $1.3$ to $0.2$, which corresponds to an impressive $84\%$ decrease.

\begin{figure}[!htb]
  \centering
  \includegraphics[width=0.7\textwidth]{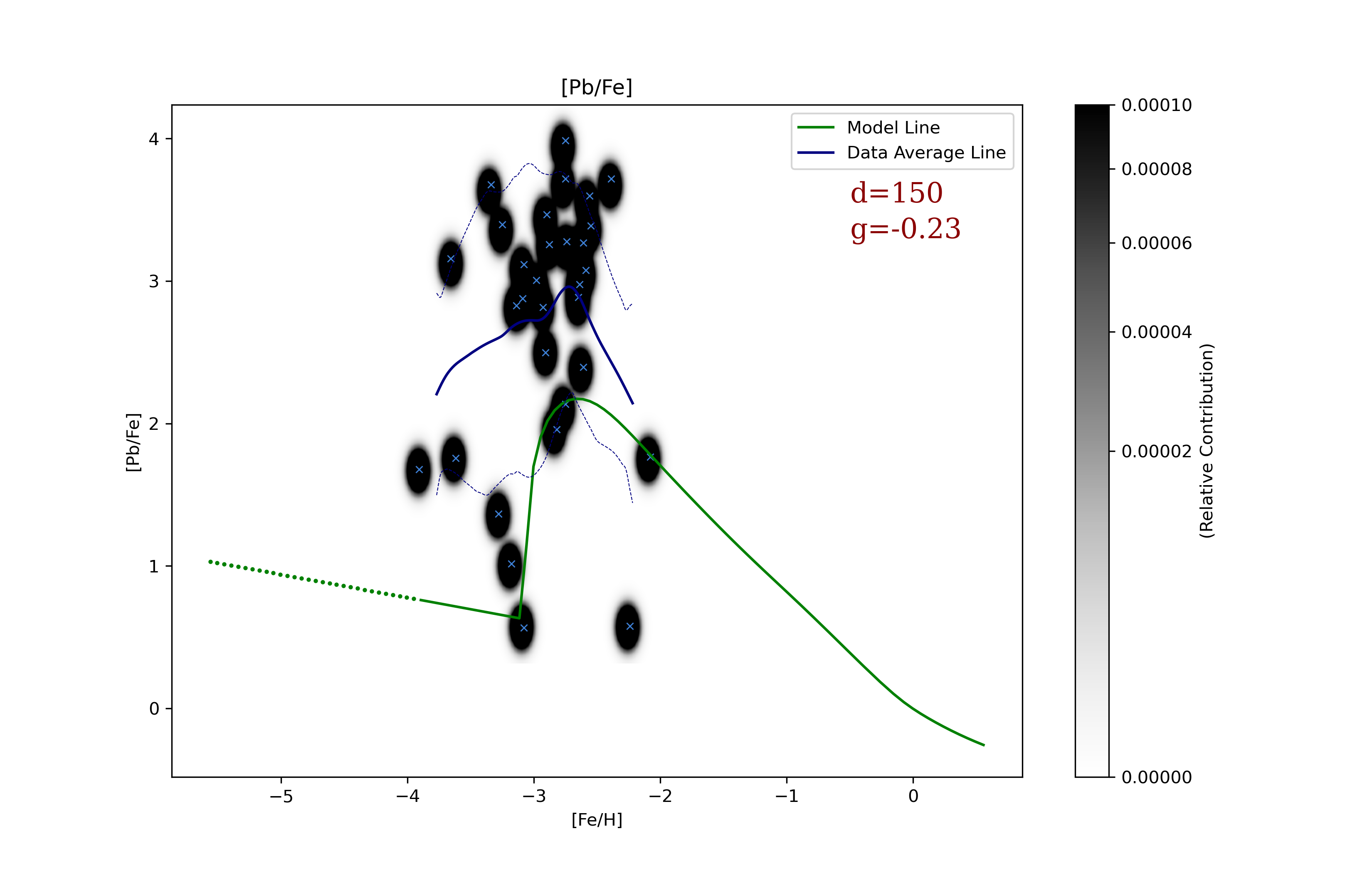}
  \caption[Published Results of Model and Data Analysis with Strong Tanh Factors (\cite{west2013metallicity})]{Published Results of Model and Data Analysis with Strong Tanh Factors (\cite{west2013metallicity}): The green line portrays the model curve incorporating a strong tanh scaling factor (150) and strong tanh shift factor (0.23). The blue line exhibits the average data trend, while the black shaded region demonstrates the uncertainty associated with individual data points. Blue crosses represent specific data points, reflecting the refined outcomes presented in this study. }
  \label{fig_3.5}
\end{figure}

\begin{figure}[!htb]
  \centering
  \includegraphics[width=0.7\textwidth]{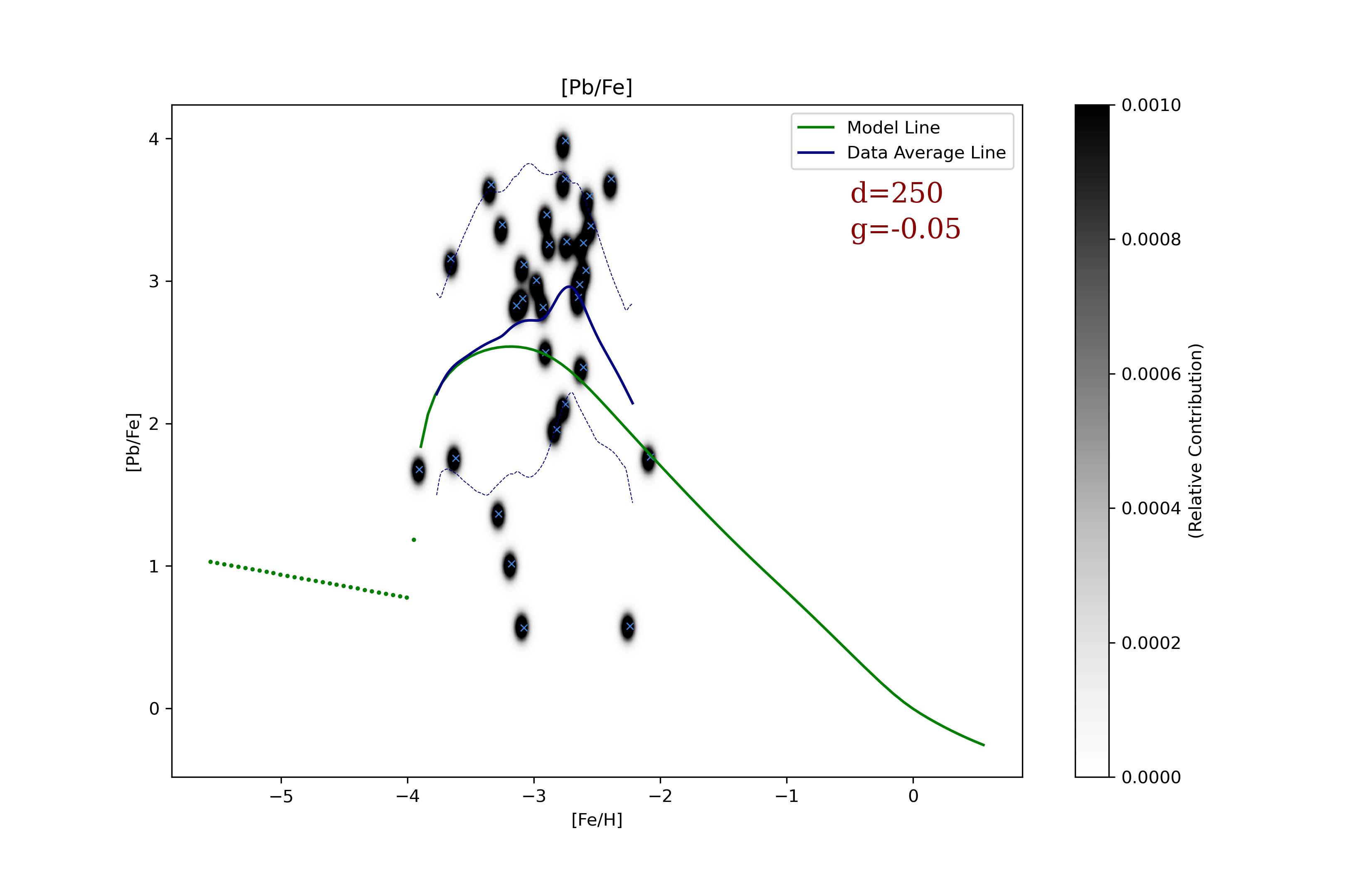}
  \caption[Enhanced Model and Data Comparison via Random Forest Algorithm]{Enhanced Model and Data Comparison via Random Forest Algorithm: The green line displays the model curve with strong tanh scaling factor (250) and strong tanh shift factor (-0.05). The blue line represents the average data trend, while the black shaded area illustrates individual data point uncertainty. Blue crosses indicate specific data points within the graph, demonstrating refined outcomes through the application of the Random Forest algorithm.}
  \label{fig_3.6}
\end{figure}

In addition to the numerical value comparisons for this particular algorithm, we also employed visual comparisons to assess the model's performance. Figures \ref{fig_3.5} and \ref{fig_3.6} compare the published results with strong tanh factors (West $\&$ Heger, 2013) and the enhanced model via the Random Forest algorithm. Both figures use a green line for the model curve, a blue line for the average data trend, a black shaded area for uncertainty, and blue crosses for specific data points. These figures highlight the improved outcomes achieved through the application of the Random Forest algorithm. Initially, we observed a significantly low and inaccurate estimation of the curve for the Lead isotope, as each data point had a substantial contribution to the model's values. The previous model appeared to place undue emphasis on the outlier data point located on the left side of the graph. However, after implementing the machine learning algorithm, we not only eliminated the outliers in the left part of the graph but also identified the unrealistic line in our model, which is the dotted green line on the left side where no data occurred, and the model is also hypothetical in that region.

The striking difference between the initial model and the current refined version highlights the algorithm's success in improving the model. Our findings demonstrate the potential of the random forest algorithm as a valuable tool for refining astrophysical models and enhancing our understanding of nucleosynthesis processes in stars.

\section{Visualizing the Isotopic Table}
After refining the astrophysical models, we incorporated the updated parameter values into our Python code to calculate and output the abundance of different reaction contributions for each of the 287 isotopes under investigation. To effectively visualize the trends in isotope reaction abundance, we varied the initial metallicity of the star from $[Z]$ ranging from $-4.5$ to $0.5$, employing a step size of $0.5$. We generated individual graphs for all isotopes and used a color-coding system to differentiate the various reactions. To facilitate easier comparison, all calculated abundances were normalized by dividing the value by the solar abundance.

The isotopic abundance table generated from our analysis will be further examined in later sections, with a focus on the trends in metallicity and the corresponding reactions that typically occur in these nucleosynthesis processes.

By visually comparing the isotopic abundance table, we can observe the relative contributions of various processes for different isotopes. In this section, we examine three distinct isotopes representing a range from light to heavy isotopes, each consisting of a variety of nuclear processes. The abundance is depicted on the y-axis using a logarithmic scale after normalization from the solar abundance, while the x-axis displays the metallicity range, which spans from $-4.5$ to $0.5$.

This in-depth analysis of isotopic abundances and their relation to metallicity provides valuable insights into the nucleosynthesis processes occurring in stars. Understanding these trends and reactions can ultimately contribute to our knowledge of stellar evolution and the formation of elements in the universe.

\subsection{$^{56}Fe$ Isotope}
\begin{figure}[!htb]
  
  \centering
  \includegraphics[width=0.8\textwidth]{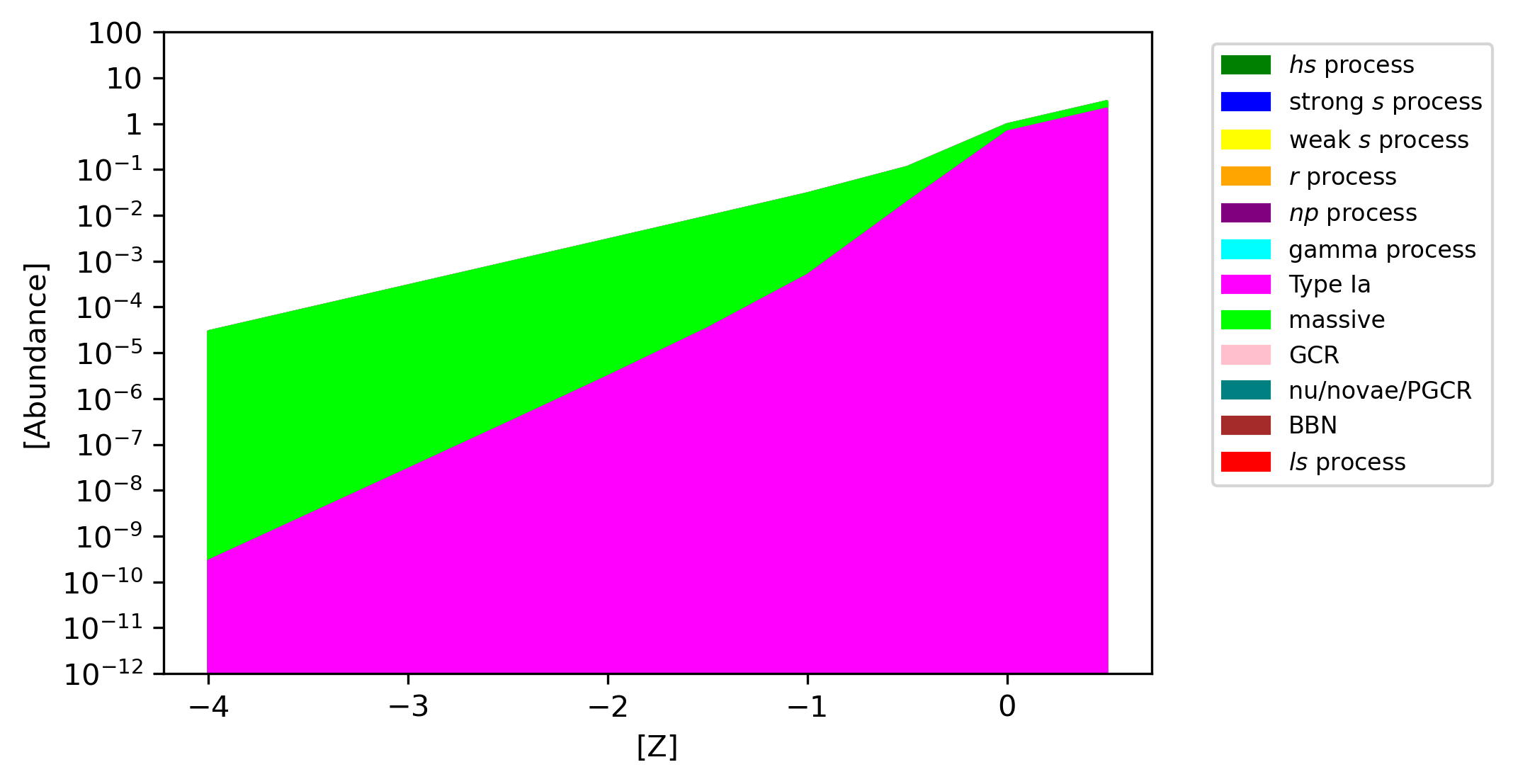}
  \caption[The abundance as a function of metallicity for $^{56}Fe$]{The abundance as a function of metallicity for $^{56}Fe$. Massive star contributions (green) and Type Ia SNe contributions (pink) are shown. This visualization permits an estimate of the relative contributions from each process to the $^{56}Fe$ abundance at any desired metallicity.\label{fig:3_7}}
\end{figure}

Iron is first formed in the late stages of massive star evolution through a series of nuclear reactions during hydrostatic burning. These abundances, however, are not returned to the interstellar medium. The iron abundance returned proceeds from explosive nucleosynthesis during the star's death. According to our model, see Figure \ref{fig:3_7}, the creation of low metallicity \(^{56}\text{Fe}\) abundances is indeed due to the results of massive star nucleosynthesis (which includes explosive nucleosynthesis and the yields from stellar winds). At low metallicities, Type Ia SNe contribute negligible amounts. When the metallicity increases, the massive star reactions contribute about \(30\%\) of the total solar abundance. At Type Ia onset, (see the change in slope in Type Ia abundances, Figure \ref{fig:3_7}), contributions increase to their solar value. At higher metallicities, massive star contributions to the iron abundance decreases, as less new iron is made relative to the initial composition.

\subsection{$^{85}Ru$ isotope}
\begin{figure}[!htb]
  \centering
  \includegraphics[width=0.8\textwidth]{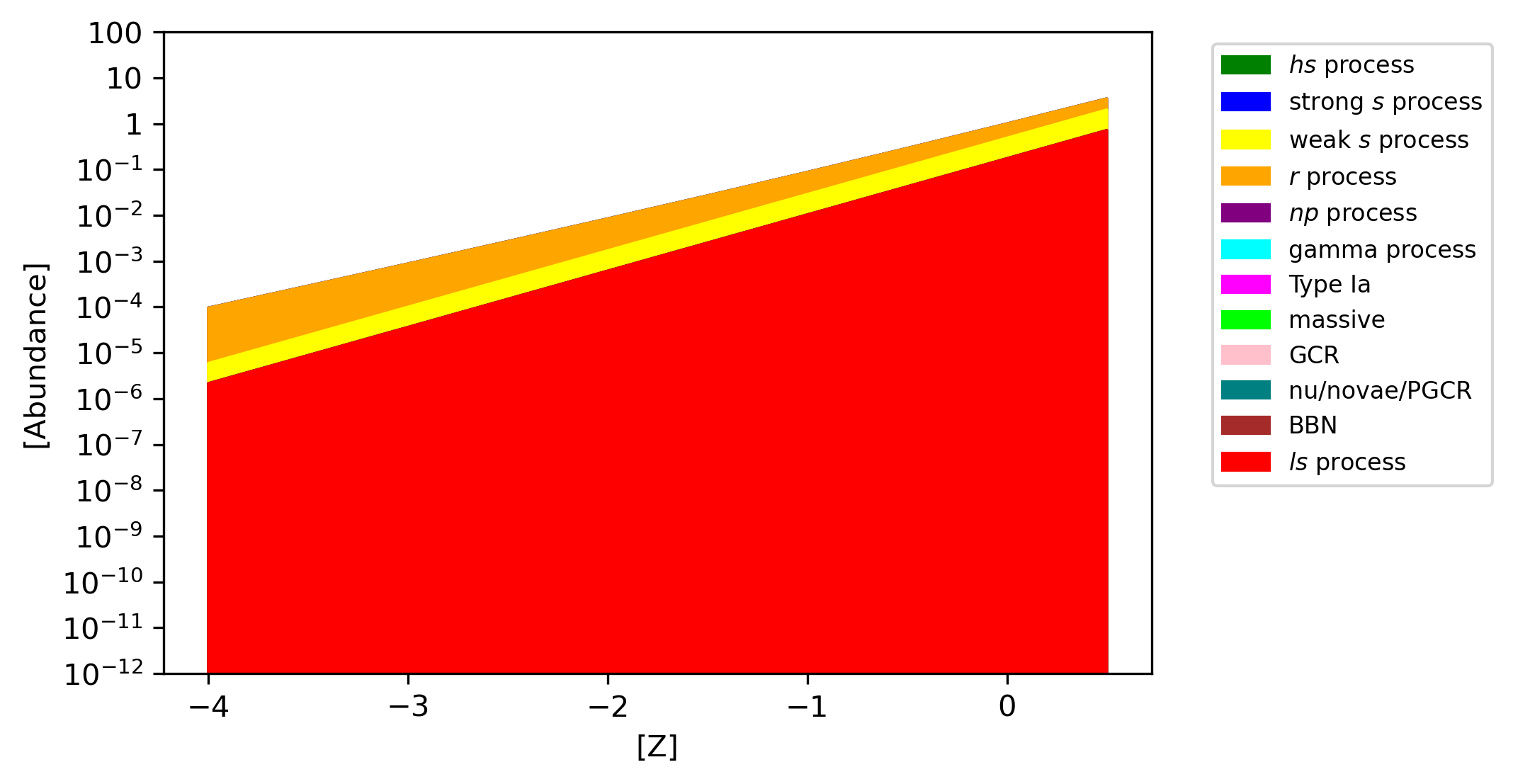}
  \caption[The abundance as a function of metallicity for $^{85}Rb$]{The abundance as a function of metallicity for $^{85}Rb$. $r$-process contributions (orange), weak $s$-process contributions (yellow) and $ls$-process contributions (red) are shown. This visualization permits an estimate of the relative contributions from each process to the $^{85}Rb$ abundance at any desired metallicity.}
  \label{fig:3_8}
\end{figure}

The Rubidium-85 isotope (see Figure \ref{fig:3_8}) is a trans-iron isotope with both primary and secondary components to its abundance. According to our model, this isotope is made from the weak \(s\)-process, \(r\)-process, and \(ls\)-process. As seen in Figure \ref{fig:3_8}, \(r\)-process contributions to this isotope dominates at all metallicities. The contribution resulting from weak \(s\)-process and \(ls\)-process obey visually distinct slopes from the \(r\)-process, due to the different fit parameters found for their scaling functions.

\subsection{$^{102}Pd$ isotope}\label{3.8}
\begin{figure}[!htb]\label{3.9}
  \centering
  \includegraphics[width=0.8\textwidth]{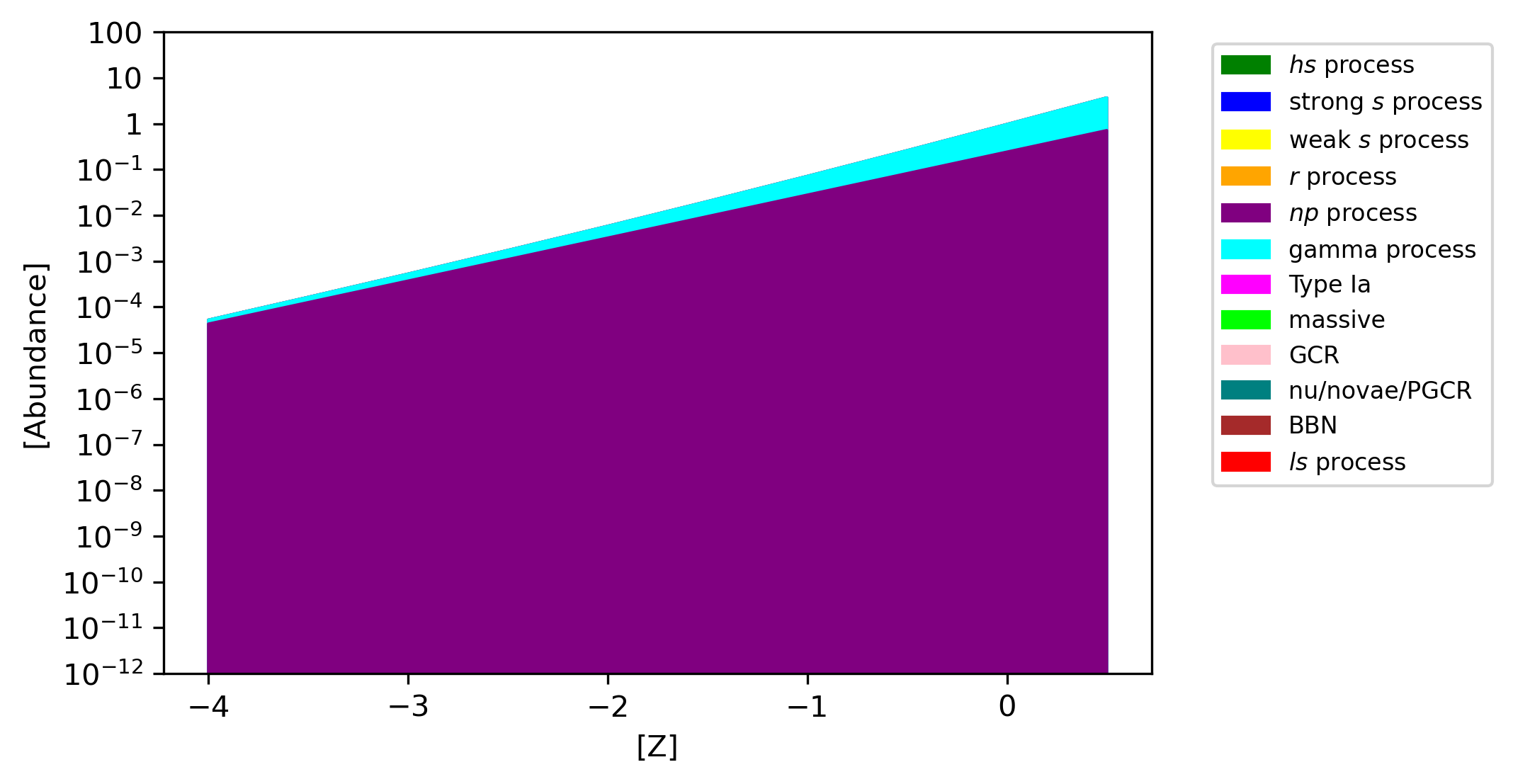}
  \caption[The abundance as a function of metallicity for $^{102}Pd$]{The abundance as a function of metallicity for $^{102}Pd$. Gamma Process contributions (cyan) and $np$ process contributions (purple) are shown. This visualization permits an estimate of the relative contributions from each process to the $^{102}Pd$ abundance at any desired metallicity.}
  \label{fig:3_9}
\end{figure}

We also highlight the scaling behavior of a Palladium isotope \ref{fig:3_9}, which, in comparison to Rubidium and iron isotopes, has contributions from different nucleosynthesis processes. Our findings indicate that the gamma process and $\nu p$-process contribute to $^{102}Pd$. However, the $\nu p$-process dominates the abundance at low metallicities due to its primary nature, whereas at higher matellicities the gamma process begins to dominate. The gamma process, or photo-disintegration events, can either have primary or secondary seed nuclei as targets. Hence, this process can either be in-between a secondary or tertiary process, although both in the present model and previous work, the parameter for it is taken as the average of primary and secondary parameters. The gamma process contributions show a tendency to increase at higher metallicities, and these different low veresus high metallicity behavior can be observed for other species, such as Strontium and Barium (See the Appendix for a complete set of all isotopic abundance scalings by process).

\section{Comparison}

In this subsection, we compare our findings with those reported in published work. The results are shown in the following table \ref{tab:importvalues}. Our analysis demonstrates a significant difference of $19.8\%$ between the published and refined results, with reduced chi-square values of 278 and 223, respectively. We attribute this improvement to the effectiveness of our machine learning methods in refining the models and adjusting certain reactions' parameters. Furthermore, visual comparisons reveal agreement for lighter isotopes but considerable differences and enhancements for heavier isotopes, corroborating our quantitative findings.

\begin{table}[!htb]\label{table1}
    \centering
    \begin{tabular}{ c c c } 
 Parameter & Best-fit Value & Description \\ 
 \hline
 $a$           &4.93(5.024)          &Type $Ia $ $tanh$ scaling factor  \\
$b$           &2.82(2.722)          &Type $Ia$ $tanh$ shifting factor  \\
$f$           &0.69(0.693)          &Fraction of solar $^{56}\text{Fe}$ from Type Ia  \\
$p$           &0.84(0.938)          &Primary process exponent  \\
$h$           &1.41(1.509)          &$hs$-process exponent  \\
$l$           &1.13(1.227)          &$ls$-process exponent  \\
$w$           &1.13(1.230)          &Weak $s$-process exponent  \\
$c$           &$-2\times 10^{-11}(-2\times 10^{-11})$         &“Strong” $tanh$ coefficient  \\
$d$           &250(150)          &“Strong” $tanh$ scaling factor  \\
$g$           &-0.05(-0.23)          &“Strong” $tanh$ shift factor  \\
 
\end{tabular}
    \caption[Found Parameter Values]{Final results after machine learning algorithms with best-fit values and comparisons with values from previous work (shown in brackets). The corresponding description for each parameter is also provided.}
    \label{tab:importvalues}
\end{table}

\section{Limitations}

Like any other study, our research has limitations that may impact the accuracy and generalizability of our findings. In this subsection, we address these limitations, which encompass various aspects of our study such as data sources, machine learning algorithms, and modeling choices.

\subsection{Data Sources}

For this project, we acquired solar isotopic abundances from \cite{lodders2020solar}. The solar abundances suffer from uncertainties in certain key isotopes, whose values are unstable across competing solar abundance data-sets (e.g., $^{14}N$). Nevertheless, the most significant uncertainties arise from spectroscopic abundance data acquired from elemental observations from stars (\cite{frebel2010stellar}). These uncertainties arise from factors such as modeling the stellar atmospheres, and these uncertainties are data-set specific and can vary across individual stars measured.

\subsection{Machine Learning Algorithm}

Our study acknowledges potential issues with the employed machine learning algorithm, which fits according to available elemental data. Whereas all data is fitted together, individual isotopic scalings may be poorly constrained if a paucity of elemental data exists for them. In essence, elements with few stellar abundance data do not receive a higher weighting considered to advance the fit. All data is considered on equal footing, which results in possibly poorly constrained fits for scalings with less data available.

The selection of the grid search algorithm for parameter tuning presents additional limitations. Given that this algorithm relies on predefined parameter values and does not extensively modify them, it may restrict the optimization process and limit the potential improvements to the model.

By acknowledging and addressing these limitations, we aim to offer a more comprehensive understanding of our study's findings and their implications. Recognizing these limitations also highlights areas where future research could further refine our understanding of astrophysical processes and improve the methodologies employed in our investigation.

\section{Uncertainty Analysis}

In this section, we will address the challenges associated with the data points and the interpretation of our findings with related to uncertainty. Due to the Gaussian assignment of the data points, the standard deviation is quite large, which leads to a significant spread in the astrophysical data. Despite this, our data model is able to track the data well and reinforce the model line averages for isotopic abundance.

As previously mentioned in the magnesium fit section, astrophysical data often suffers from a large spread, as demonstrated by the red line which represents the data line with Gaussian assignments. This spread results in several fit parameters yielding reduced chi-square values of less than 1. To mitigate this issue, we have introduced a scaling function for the astrophysical processes derived from nuclear physics. This approach ensures that our model selects the best average data without overfitting.

However, it is important to note that due to the inherent uncertainties in astrophysical data, our model provides only average chemical abundance estimates and does not predict any particular isotopic abundance in the interstellar medium. This limitation should be taken into account when interpreting the results of this study and applying them to specific astrophysical contexts.

\section{Conclusion}\label{Ch5}
In this project, our primary objective was to refine and enhance the existing construction of Galactic isotopic decomposition for all stable isotopes, by leveraging advanced machine learning algorithms. Understanding the astrophysical processes responsible for isotope synthesis is crucial for gaining insights into the origins of elements in the universe and improving our knowledge of stellar nucleosynthesis processes. The scope of our research encompassed the development of a comprehensive isotopic table, representing the relationship between a star's initial metallicity and isotopic abundance. To achieve this, we examined various astrophysical processes responsible for isotope synthesis using solar abundance data. We successfully modified the model, achieving lower reduced chi-square values compared to the published results by West et al. (2013). This work laid the foundation for isotopic decomposition as a function of metallicity, based on elemental observational data and underlying nucleosynthesis processes in complex environments like the Galaxy. Our key scientific findings include: Verification and refinement of light isotope parameter values: Through the application of machine learning algorithms, we not only verified the previous work on parameter values but also provided a more detailed analysis of fitting lines for specific groups of light isotopes.Significant improvement in heavy isotope predictions: By utilizing a larger dataset to train the model and replacing the simple estimation methods from previous results, we achieved a substantial improvement in the predictions for heavy isotopes.The project's broader implications involve providing isotopic abundances that can be used as initial abundances for stellar models in future work or other nucleosynthesis studies. Furthermore, our research contributes to the improvement of input isotopic abundance patterns by fitting a complex model structure with interconnected parameters, marking the first time this has been done in a systematic manner.Looking ahead, we propose several future extensions of this project: Integration with Stellar Evolution Software: Implementing the refined model into a realistic simulation of stellar evolution software, such as MESA, will allow us to analyze the impact of initial metallicity on each phase of a star's evolution and isotope output, providing invaluable insights into the life cycle of stars.Expanding the Application of Machine Learning Algorithms: We suggest broadening the application of the Random Forest Algorithm to optimize more parameter fittings and exploring advanced neural network machine learning techniques, such as deep learning, for further improvements in model accuracy. Validation and Comparison of Isotopic Scaling Models: Validating the success of the machine learning methods by applying different isotopic scaling models will enable us to determine the most accurate model for predicting isotopic abundances. Additionally, this will help enhance principle component analysis based on physical nucleosynthesis processes, furthering our understanding of stellar nucleosynthesis.By pursuing these future research directions, we can continue to improve our understanding of isotopic decomposition in relation to stellar metallicity and advance the development of more accurate and comprehensive stellar models and nucleosynthesis studies. This honors thesis serves as an essential stepping stone towards a deeper understanding of the origins and evolution of elements in the universe.

\bibliographystyle{unsrt}  


\end{document}